\documentclass[letterpaper,twocolumn,aps,prl,showkeys,nofootinbib]{revtex4-1}
\usepackage[T1]{fontenc}
\usepackage[latin9]{inputenc}
\usepackage{color}
\usepackage{verbatim}
\usepackage{mathtools}
\usepackage{amsmath}
\usepackage{amssymb}
\usepackage{graphicx}
\usepackage[unicode=true,
 bookmarks=false,
 breaklinks=false,pdfborder={0 0 1},backref=false,colorlinks=true]
 {hyperref}
\hypersetup{
 linkcolor=blue, citecolor=blue}
\usepackage{breakurl}

\makeatletter

\usepackage{microtype}
\usepackage{stix}

\makeatother

\begin{document}

\title{Demonstration of efficient nonreciprocity in a microwave optomechanical
circuit}

\author{G. A. Peterson, F. Lecocq, K. Cicak, R. W. Simmonds, J. Aumentado,
and J. D. Teufel}

\affiliation{National Institute of Standards and Technology, 325 Broadway, Boulder,
CO 80305, USA}
\begin{abstract}
The ability to engineer nonreciprocal interactions is an essential
tool in modern communication technology as well as a powerful resource
for building quantum networks. Aside from large reverse isolation,
a nonreciprocal device suitable for applications must also have high
efficiency (low insertion loss) and low output noise. Recent theoretical
and experimental studies have shown that nonreciprocal behavior can
be achieved in optomechanical systems, but performance in these last
two attributes has been limited. Here we demonstrate an efficient,
frequency-converting microwave isolator based on the optomechanical
interactions between electromagnetic fields and a mechanically compliant
vacuum gap capacitor. We achieve simultaneous reverse isolation of
more than 20~dB and insertion loss less than 1.5~dB over a bandwidth
of 5~kHz. We characterize the nonreciprocal noise performance of
the device, observing that the residual thermal noise from the mechanical
environments is routed solely to the input of the isolator. Our measurements
show quantitative agreement with a general coupled-mode theory. Unlike
conventional isolators and circulators, these compact nonreciprocal
devices do not require a static magnetic field, and they allow for
dynamic control of the direction of isolation. With these advantages,
similar devices could enable programmable, high-efficiency connections
between disparate nodes of quantum networks, even efficiently bridging
the microwave and optical domains.
\end{abstract}

\date{\today}


\maketitle

\begin{figure}
\includegraphics[width=1\columnwidth]{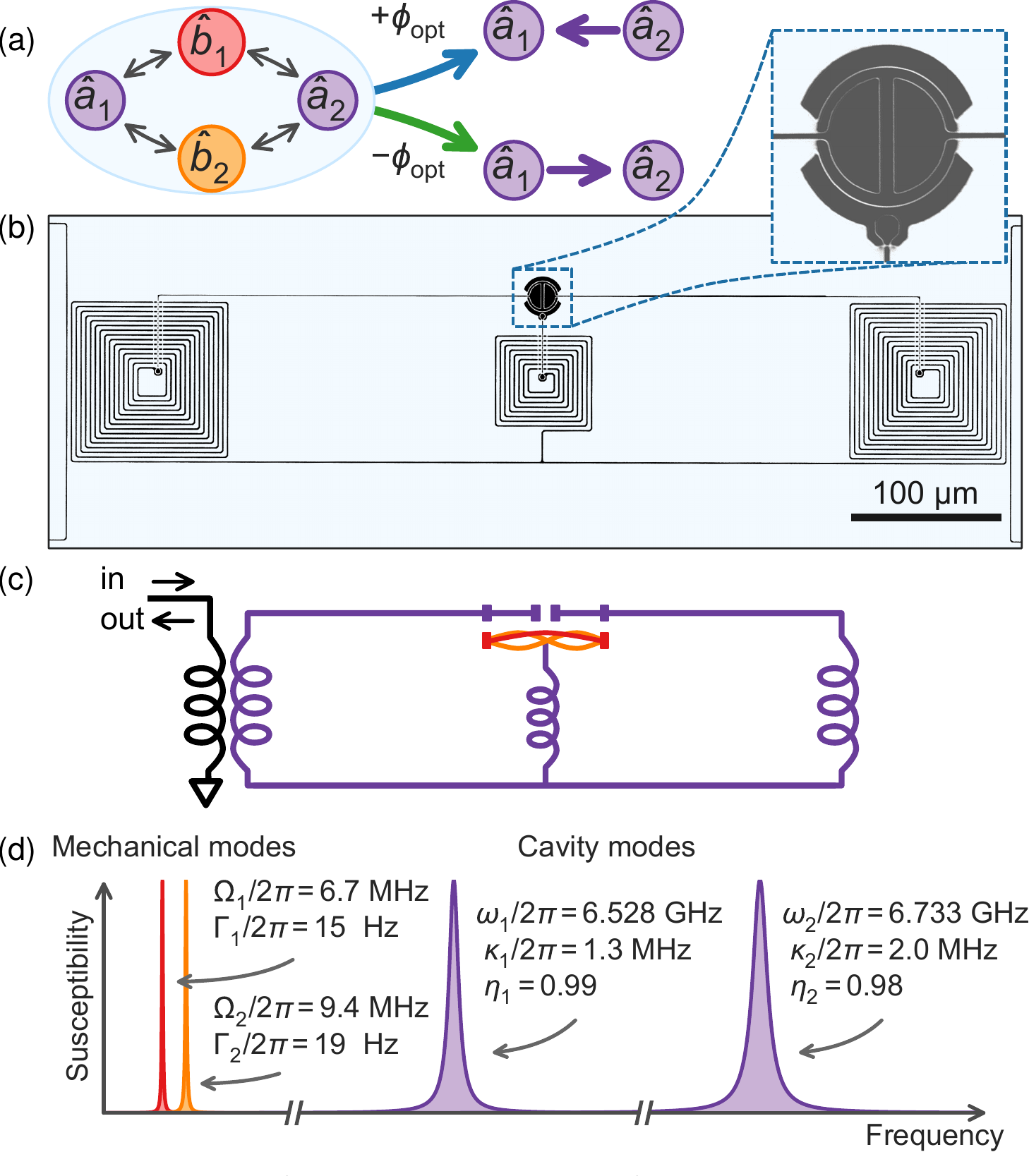} \protect\caption{Concept and experimental realization. (a) Mode-coupling diagrams for
the optomechanical isolator. Optomechanical interactions (double-sided
arrows) between two cavity modes ($\hat{a}_{1}$ and $\hat{a}_{2}$)
and two mechanical modes ($\hat{b}_{1}$ and $\hat{b}_{2}$) induce
directional scattering between the two cavities when the parametric
loop phase is equal to its optimal values $\pm\phi_{\text{opt}}$.
(b) Optical microscope image of the device. A microfabricated vacuum-gap
capacitor (inset) resonates with spiral inductors to produce two electromagnetic
cavities. (c) Schematic of the optomechanical circuit. Input signals
from microwave generators couple inductively to the device and reflect
back through the amplification chain to be measured by a network or
spectrum analyzer. (d) Frequency space diagram. Mode susceptibilities
are plotted versus frequency. Two mechanical modes and two cavity
modes are characterized by their resonant frequencies ($\Omega_{k}$
and $\omega_{j}$) and their linewidths ($\Gamma_{k}$ and $\kappa_{j}$),
and the cavities are further characterized by their coupling efficiencies
$\eta_{j}$.}
\label{fig:diagram} 
\end{figure}

Many branches of physics and engineering employ nonreciprocal devices
to route signals along desired paths of measurement networks. Conceptually,
the simplest nonreciprocal element is the isolator, a two-port device
that transmits signals from the first to the second port but strongly
attenuates in the reverse direction \cite{JalasPetrovEichEtAl2013}.
Placing an ideal isolator (or its close relative, the circulator)
between two systems allows the first system to influence the second
but not vice versa. This nonreciprocal functionality enables, for
example, telecommunication antennas to transmit and receive signals
at the same time. Another example relevant for future applications
is quantum signal processing, where the strict demands of quantum
measurement require isolators with high performance in several metrics,
including not only large isolation, but also high efficiency and low
noise \cite{Kimble2008}.

Well-established technology uses magnetic materials to achieve nonreciprocity
for both microwave and optical frequencies \cite{Polder1949,Hogan1952,ApletCarson1964}.
While these conventional devices have enabled much of the progress
in classical and quantum signal processing, overcoming their limitations
could lead to exciting new developments in both areas. For example,
these components are typically bulky, not chip-compatible, and because
they require strong magnetic fields, are incompatible with superconducting
technology. Signal losses due to these conventional nonreciprocal
devices have now become the bottleneck for the overall efficiency
of, for example, state-of-the-art microwave measurements \cite{RochFlurinNguyenEtAl2012,KindelSchroerLehnert2016,ClarkLecocqSimmondsEtAl2016}.

In recent years, there has been interest in developing nonmagnetic
nonreciprocal devices to replace conventional isolators and overcome
the limitations discussed above for superconducting microwave applications
\cite{EstepSounasSoricEtAl2014,KamalRoyClarkeEtAl2014,AbdoSliwaShankarEtAl2014,KerckhoffLalumiereChapmanEtAl2015,SliwaHatridgeNarlaEtAl2015,MacklinOextquotesingleBrienHoverEtAl2015,MetelmannClerk2015}
as well as limitations that arise in optical and room temperature
isolation \cite{KangButschRussell2011}. Schemes based on coupled-mode
physics can break reciprocity without a static magnetic field if the
coupling is parametrically modulated in time \cite{JalasPetrovEichEtAl2013}.
Producing isolation further requires the coherent interference of
two paths from one port to another as well a reservoir to absorb the
backward-propagating power \cite{MetelmannClerk2015,Ranzani2015}.
These schemes are particularly promising because they can naturally
integrate with existing chip-based superconducting technology \cite{Ranzani2015,AbdoSliwaFrunzioEtAl2013,AbdoSliwaShankarEtAl2014,KamalClarkeDevoret2011,KamalRoyClarkeEtAl2014,LecocqRanzani2017}.

One route for efficient parametric nonreciprocity in the microwave
domain is to use Josephson junctions to couple superconducting circuits
\cite{SliwaHatridgeNarlaEtAl2015,LecocqRanzani2017}. More recently,
theoretical proposals \cite{HafeziRabl2012,Ranzani2015,MetelmannClerk2015,XuLiChenEtAl2016}
and experiments \cite{ShenZhangChenEtAl2016,RuesinkMiriAluEtAl2016,FangLuoMetelmannEtAl2017}
have begun exploring the parametric coupling between an electromagnetic
cavity and a mechanical oscillator as an alternative mode-coupling
mechanism for nonreciprocity. These optomechanical systems are attractive
because of their wide applicability beyond microwave frequencies and
cryogenic environments. For example, efficient, reciprocal frequency
conversion using optomechanics has already been demonstrated in both
the microwave \cite{LecocqClarkSimmondsEtAl2016} and optical \cite{LiuDavancoAksyukEtAl2013,LiDavancoSrinivasan2016}
frequency bands, as well as in conversion between the two \cite{AndrewsLehnert2014}.
Nonreciprocal optomechanical devices, however, have yet to show the
efficiencies and noise properties needed for most applications.

Combining two independent optomechanical frequency converters gives
a natural way to achieve the interference needed for nonreciprocity.
Here we realize this interference by simultaneously coupling two electromagnetic
cavity modes to two distinct vibrational modes of a mechanical membrane.
We illustrate this concept for achieving nonreciprocal frequency conversion
between the two cavities in Fig.~\ref{fig:diagram}(a).

To understand the optomechanical isolator, we begin with the fundamental
parametric interaction between an electromagnetic cavity and a mechanical
oscillator \cite{AspelmeyerKippenbergMarquardt2014}. A general multimode
cavity optomechanical system consists of a set of cavity resonances
and mechanical modes. Consider a given cavity mode $j$ with resonant
frequency $\omega_{j}$ and linewidth $\kappa_{j}$ and a given mechanical
mode $k$ with frequency $\Omega_{k}$ and intrinsic linewidth $\Gamma_{k}$,
obeying $\Gamma_{k}\ll\kappa_{j}<\Omega_{k}\ll\omega_{j}$. The position
of the mechanical oscillator tunes the cavity frequency, providing
the mechanism of coupling. Analysis of the equations of motion for
the cavity and mechanical mode annihilation operators, $\hat{a}_{j}$
and $\hat{b}_{k}$, shows that a strong electromagnetic field (the
\emph{drive}) applied at a frequency near the red sideband (defined
by $\omega_{jk}=\omega_{j}-\Omega_{k}$) induces an effective beam-splitter
interaction. The interaction Hamiltonian is $\hbar(g_{jk}\hat{a}_{j}\hat{b}_{k}^{\dagger}+g_{jk}^{*}\hat{a}_{j}^{\dagger}\hat{b}_{k})$,
where $\hbar$ is the reduced Planck constant, and the coupling rate
$g_{jk}$ is a complex number with phase and amplitude set by the
drive. We parameterize the coupling strength in terms of the cooperativity
$C_{jk}=4|g_{jk}|^{2}/(\kappa_{j}\Gamma_{k})$.

Our optomechanical isolator is fully described by the general theory
of linear coupled-mode systems \cite{Louisell1960,Ranzani2015,LecocqRanzani2017}.
In the quantum input-output formalism \cite{Gardiner1985}, each mode
$\hat{a}_{j}$ couples to its environmental input and output operators
$\hat{a}_{j,\text{in}}$ and $\hat{a}_{j,\text{out}}$ through the
standard input-output boundary conditions. The scattering matrix elements
are defined as the ratios of output to input field amplitudes, $S_{jk}=\left\langle \hat{a}_{j,\text{out}}\right\rangle /\left\langle \hat{a}_{k,\text{in}}\right\rangle $,
where $\left\langle \cdot\right\rangle $ indicates expectation value.
Demonstrating an efficient isolator requires maximizing the forward
transmission $|S_{jk}|^{2}$ while minimizing the reverse transmission
$|S_{kj}|^{2}$.

We experimentally create a system consisting of two cavity modes and
two mechanical modes by designing and fabricating a superconducting
circuit of aluminum on a sapphire substrate \cite{CicakLiStrongEtAl2010,TeufelLiAllmanEtAl2011,LecocqTeufelAumentadoEtAl2015},
as shown and characterized in Fig.~\ref{fig:diagram}(b--d). A vacuum-gap
capacitor combined with an inductive network defines two microwave
cavities with resonant frequencies $\omega_{1}/2\pi=6.528$~GHz and
$\omega_{2}/2\pi=6.733$~GHz and linewidths $\kappa_{1}/2\pi=1.3$~MHz
and $\kappa_{2}/2\pi=2.0$~MHz. We design the cavities to be highly
overcoupled so that the intentional inductive coupling rate to the
measurement line, $\kappa_{\text{ext}}$, dominates the total dissipation
rate of each cavity, $\kappa_{\text{tot}}$. The coupling efficiencies
for each cavity, defined as $\eta_{j}\equiv\kappa_{j,\text{ext}}/\kappa_{j,\text{tot}}$,
are measured to be $\eta_{1}\simeq0.99$ and $\eta_{2}\simeq0.98$.
The vacuum-gap capacitor has a mechanically compliant top plate that
vibrates with several spectrally distinct mode frequencies. In this
experiment, we use the two lowest-frequency vibrational modes at $\Omega_{1}/2\pi=6.7$~MHz
and $\Omega_{2}/2\pi=9.4$~MHz with intrinsic linewidths $\Gamma_{1}/2\pi=15$~Hz
and $\Gamma_{2}/2\pi=19$~Hz, as determined by independent measurements
of the energy dissipation rate. We place the device in a dilution
cryostat with a base temperature of 19~mK and interrogate the circuit
with signals routed from microwave generators and a vector network
analyzer. From room temperature components, input signals pass through
attenuators, reflect off the device at a circulator, and pass through
a cryogenic high-electron-mobility transistor amplifier, with more
amplification at room temperature. We operate the device as a single
physical port measured in reflection; ports 1 and 2 used hereafter
refer to input or output signals near the resonant frequencies of
cavities 1 and 2.

\begin{figure}[t]
\includegraphics[width=1\linewidth]{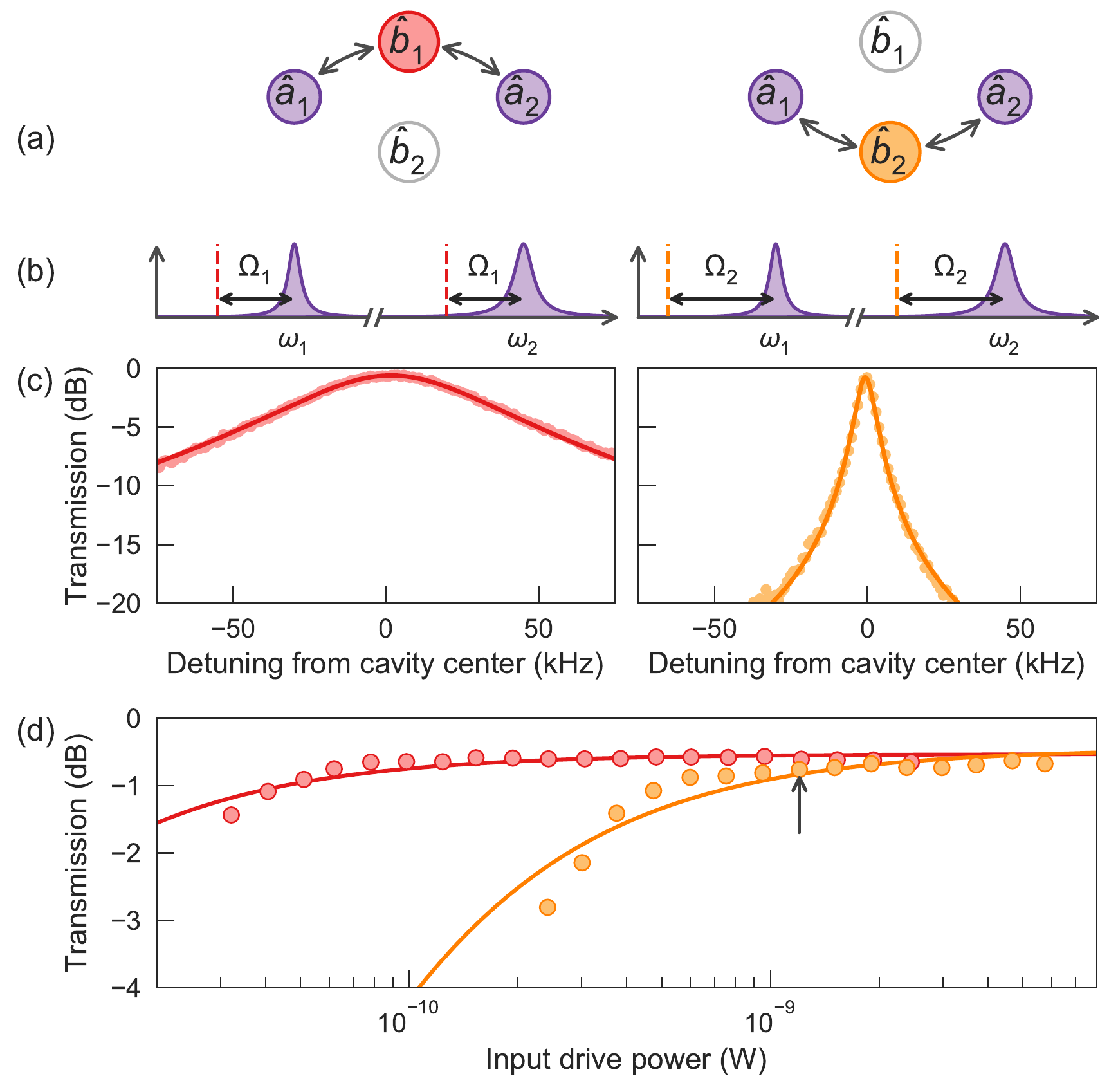} \protect\caption{Reciprocal mechanically-mediated frequency conversion. (a) Mode-connection
diagrams. Double-sided arrows indicate driven optomechanical interactions.
(b) Frequency space diagrams. A red-detuned drive applied at each
cavity induces frequency conversion through one mechanical mode. Dashed
lines indicate frequencies of microwave drives. (c) Measured magnitude
of reciprocal transmission from cavity 2 to cavity 1 as a function
of the probe detuning from cavity center for a particular drive power.
Frequency conversion through the first mechanical mode is shown in
red on the left and through the second mechanical mode in orange on
the right. Solid lines are fits to Lorentzian lineshapes. (d) Maximum
transmission as a function of total input drive power for the first
(red) and second (orange) mechanical modes. Solid lines are fit to
a model described in \cite{LecocqClarkSimmondsEtAl2016}. The arrow
indicates the drive power used in (c).}
\label{fig:FC} 
\end{figure}

As reciprocal frequency conversion forms the basis for the optomechanical
isolator, we first demonstrate this process through each mechanical
mode (Fig.~\ref{fig:FC}). In this scheme, one microwave drive is
applied at each cavity's red sideband with respect to a single mechanical
mode; a signal entering one cavity down-converts to the mechanical
mode and then up-converts to the other cavity (Fig.~\ref{fig:FC}a,b). 

In Fig.~\ref{fig:FC}(c) we show the reciprocal transmission from
one cavity to the other as a function of detuning from the cavity
center frequencies. We calibrate the scattering parameters using methods
described previously \cite{AndrewsLehnert2014,LecocqClarkSimmondsEtAl2016}.
A drive power of approximately 1~nW damps mechanical mode 1 (left)
to about 70 kHz and mode 2 (right) to 7 kHz. These damping rates are
comparable to those used later in the nonreciprocal scheme. We achieve
transmission above $-0.6$~dB through each mode, limited by cavity
loss and pump strength imbalance. At our highest drive powers, the
bandwidths of frequency conversion through the mechanical modes reach
150~kHz and 35~kHz. Our frequency converter operates in the high
cooperativity limit, as evidenced by the large ratios of damped mechanical
linewidths to intrinsic linewidths and the plateau in transmission
versus input power, shown in Fig.~\ref{fig:FC}(c).

\begin{figure}
\includegraphics[width=1\linewidth]{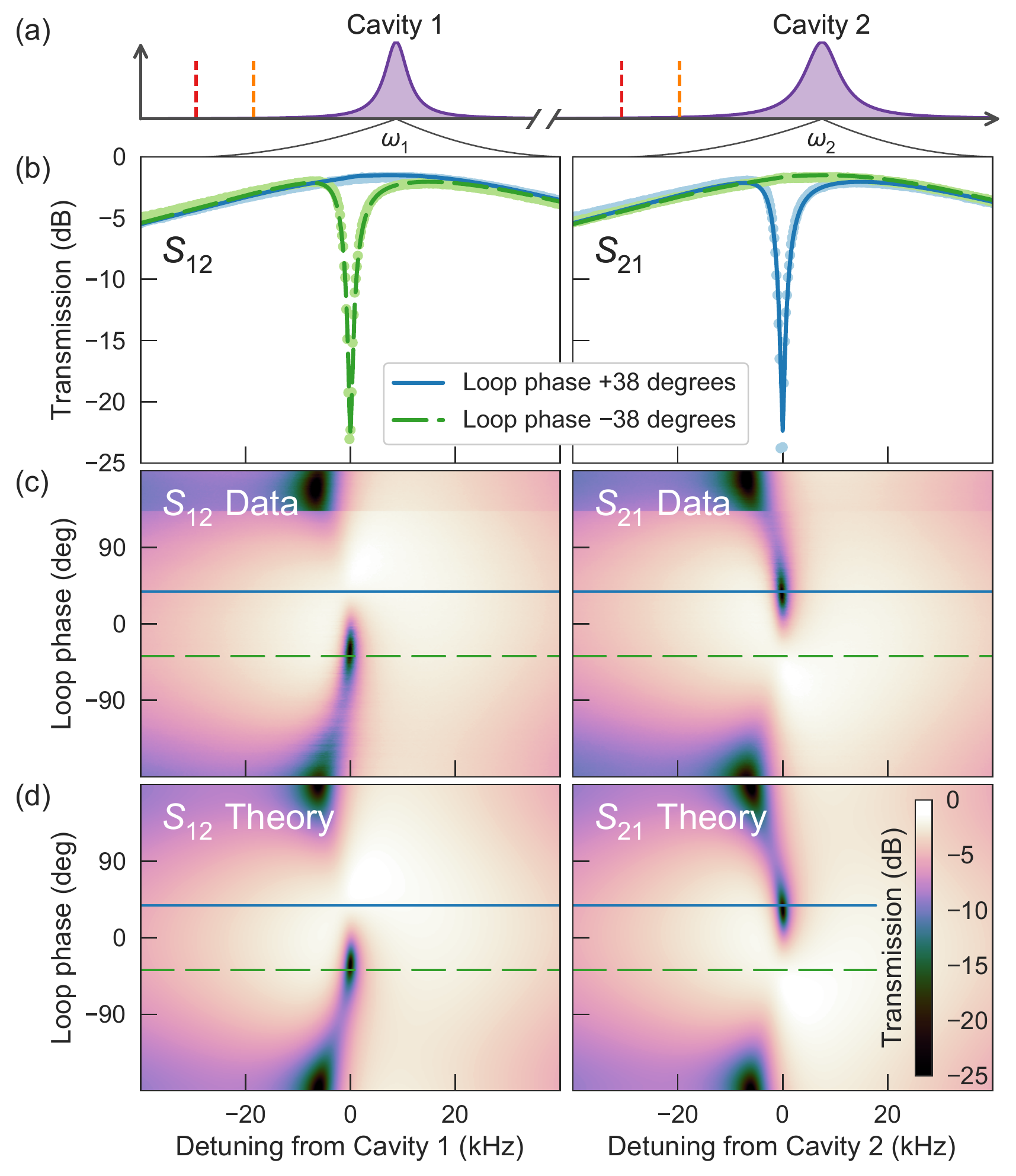} \protect\caption{Optomechanical isolation. (a) Frequency space diagram. Four drives
(dashed lines) induce frequency conversion between the two cavities
through both mechanical modes simultaneously. (b) Measured magnitude
of transmitted signal received at cavity 1 (left) and cavity 2 (right)
for two choices of loop phase. At $\phi=+38$ degrees (solid blue)
signals are transmitted from cavity 2 to cavity 1 and attenuated in
the reverse direction. The behavior reverses at $\phi=-38$ (dashed
green). Solid lines are fits to the expanded coupled-mode theory model
described in the text. (c) Transmission (colorscale) as a function
of detuning and loop phase. Lines show the source of data shown in
panel (b). (d) Result of the least-squares fit of the two-dimensional
data in (c).}
\label{fig:nonrec} 
\end{figure}

Now, to realize the optomechanical isolator, we drive two branches
of mechanically-mediated frequency conversion simultaneously. Figure~\ref{fig:nonrec}(a)
shows the frequency space diagram of the experiment, with dashed lines
indicating the frequencies of the four drives. Ideal isolation maximizes
the magnitude of the transmission difference, defined as $\Delta T=|S_{21}|^{2}-|S_{12}|^{2}$.
Transmission difference lies between $-1$ and $1$, making it a useful
metric because it simultaneously favors high reverse isolation and
low insertion loss, both important for quantum signals applications.

To achieve ideal isolation at the cavity resonances, the powers, frequencies,
and relative phases of the four drives must be tuned to optimal values.
Assuming the cavity linewidths are much larger than the mechanical
mode linewidths and the optomechanical cooperativities are large,
we can derive closed-form solutions for the optimal drive parameters
and the scattering matrix by analytically maximizing the function
$\Delta T$ (see supplementary information). First, the drive powers
should be such that the cooperativities for all four optomechanical
couplings are equal (let their shared value be $C$). Isolation performance
increases with this cooperativity as $\Delta T=\eta_{1}\eta_{2}(1-(2C)^{-1})$.
The second condition sets the drive frequencies. One might expect
that tuning the four drives to the exact red sideband frequencies
would be ideal. In fact, this configuration leads to reciprocal behavior
precisely at the cavity center frequencies. Permitting detuning of
the drive pairs from the red sidebands allows nonreciprocal transmission
to occur on resonance with the cavities. The optimal drive detunings
are $\delta_{j}=\pm(-1)^{j}\Gamma_{j}\sqrt{2C-1}/2$, where $\delta_{j}$
is the detuning from the red sideband of the drives that connect to
the $j$th mechanical mode. The third important condition relates
to the optimal relative drive phases. A signal traversing the loop
in mode space acquires a phase, called the \textit{loop phase} $\phi$.
Because the frequency conversion processes are parametric, this phase
is related to the sum of the relative phases of the four drives, making
it a dynamically tunable parameter. Under the assumptions mentioned
above, the optimal values of the loop phase are $\phi_{\text{opt}}=\pm\arccos(1-1/C)$.

After substituting these optimized drive parameters, and further letting
$\eta_{1}=\eta_{2}=1$ and taking the large $C$ limit, the full scattering
matrix becomes 
\begin{equation}
|\mathbf{S}|^{2}=\left(\begin{array}{cccc}
0 & 0 & 1/2 & 1/2\\
1 & 0 & 0 & 0\\
0 & 1/2 & 1/4 & 1/4\\
0 & 1/2 & 1/4 & 1/4
\end{array}\right),\label{eq:Smatrix}
\end{equation}
where the mode basis is ordered $(\hat{a}_{1},\hat{a}_{2},\hat{b}_{1},\hat{b}_{2})$.
We see that the upper left corner defines the ideal $2\times2$ isolator,
perfectly isolating cavity 1 from cavity 2. The other matrix elements
describe scattering of signals input to the mechanical modes. At the
opposite loop phase, the scattering matrix becomes the transpose of
that shown above, isolating cavity 2 from cavity 1.

In contrast to reciprocal frequency conversion, the mechanical dissipation
plays a key role in the nonreciprocal behavior of the device. This
is a consequence of power conservation; isolation can only occur if
power entering a cavity mode can be completely routed into the mechanical
environments. The mechanical modes are coupled to their environment
with fixed rates $\Gamma_{j}$. So, while the bandwidth $\Gamma_{R}$
of reciprocal mechanically-mediated frequency conversion increases
with cooperativity as $\Gamma_{R}=\Gamma_{j}(1+2C)$ \cite{LecocqClarkSimmondsEtAl2016},
the nonreciprocal bandwidth $\Gamma_{NR}$ for the isolating system
in the high-cooperativity limit is $\Gamma_{NR}=4\Gamma_{1}\Gamma_{2}/(\Gamma_{1}+\Gamma_{2})$,
involving only the intrinsic mechanical linewidths, independent of
cooperativity. As we explore below, damping processes that occur outside
the nonreciprocal loop produce effective mechanical linewidths and
therefore allow the nonreciprocal bandwidth to increase.

Before describing the data, it is necessary to include an important
deviation of our device from the simple system of four modes described
thus far. Ideally, a given parametric drive couples a single mechanical
mode to a single cavity mode. In practice, however, this drive also
couples the other mechanical mode to the cavity off-resonantly. This
residual coupling damps and cools the mechanical modes. These effects
can be rigorously accounted for in the coupled equations of motion
by expanding the mode basis to include all interactions (see supplementary
information). Modeling these processes as additional modes allows
us to accurately map the experimental system to the simpler system
of four modes with effective mechanical linewidths and effective cooperativities.
By damping the mechanical modes to widths much larger than the intrinsic
mechanical linewidths, these off-resonant terms greatly enhance the
bandwidth and noise performance of the isolator, but they also reduce
the effective cooperativities attainable. Modeling the extra damping
terms gives us a predictive theory with which to tune the device and
arrive at ideal performance parameters.

Figure~\ref{fig:nonrec}(b) shows the measured transmission from
cavity 2 to cavity 1 (left) and from 1 to 2 (right) at two loop phases
for a particular drive configuration found from the tuning process.
On cavity resonance at $\phi=+38$~degrees (solid blue), we see high
transmission (insertion loss of $1.5$~dB) from cavity 2 to cavity
1 but low transmission (isolation of $21$~dB) from cavity 1 to 2
over a frequency band of $5$~kHz. At $\phi=-38$~degrees (dashed
green), the behavior reverses. We collect data at many loop phases,
shown in Fig.~\ref{fig:nonrec}(c) with horizontal lines indicating
the cuts shown in Fig.~\ref{fig:nonrec}(b). We fit the data to the
expanded coupled-mode model using a two-dimensional nonlinear least-squares
fit, the result of which is shown in Fig.~\ref{fig:nonrec}(d), demonstrating
excellent agreement with the data. Mapping our expanded model onto
the four-mode system gives the effective system parameters. The effective
mechanical linewidths are $\Gamma_{1,\text{eff}}/2\pi=1.6$~kHz and
$\Gamma_{2,\text{eff}}/2\pi=7.5$~kHz, in agreement with the nonreciprocal
bandwidth of 5~kHz. The four effective cooperativities are $(C_{11},C_{12},C_{21},C_{22})=(5.4,5.7,2.9,2.0)$,
where the notation $C_{jk}$ indicates the cooperativity coupling
cavity $j$ to effective mechanical mode $k$.

While the loop phase of $\phi=\pm38$~degrees gives good balance
between the goals of high reverse isolation and low insertion loss,
other loop phases can be maximize these metrics individually. For
the drive configuration shown here, the insertion loss can be as low
as 1.16~dB ($\approx77\%$~efficiency) at $\phi=\pm63$~degrees
at the expense of reducing reverse isolation to 9.2~dB. Alternatively,
the reverse isolation can be tuned arbitrarily high near $\phi=\pm32$~degrees
at the expense of slightly increasing the insertion loss. In our system,
we observe isolation at a single frequency as high 49~dB with corresponding
insertion loss of 1.9~dB.

\begin{figure}
\includegraphics[width=1\linewidth]{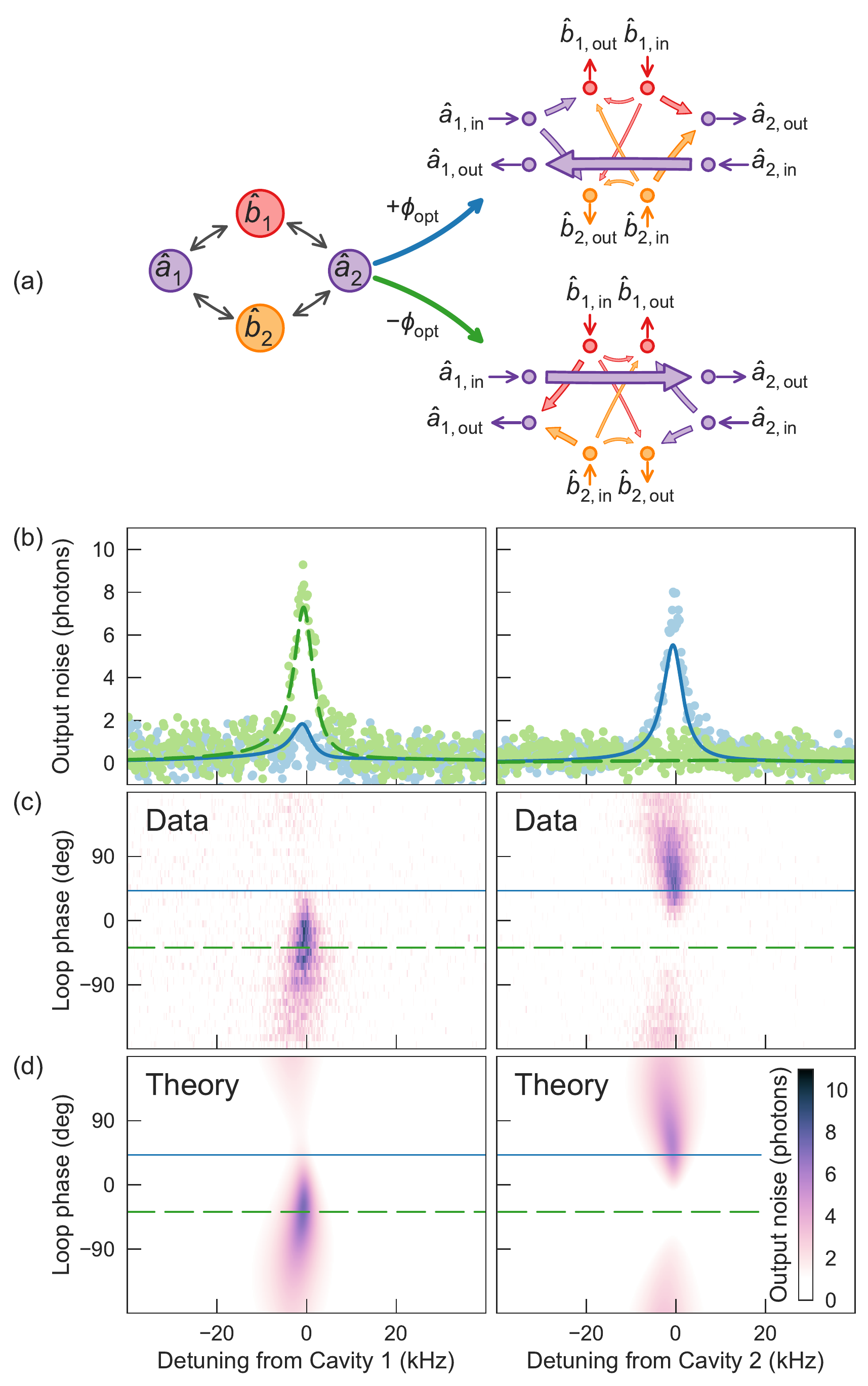} \protect\caption{Noise performance of the optomechanical isolator. (a) Graphical representation
of signal flow. The mode-connection diagram (left) induces signal
flow diagrams (right) at the optimal loop phases $\pm\phi_{\text{opt}}$.
Arrow widths are proportional to their corresponding scattering matrix
element (Eq.~\ref{eq:Smatrix}). (b) Measured output noise at cavities
1 (left) and 2 (right) near loop phases $+38$ degrees (solid blue)
and $-38$ degrees (dashed green). We have subtracted the noise offsets
of 31.5 and 22.8 photons due to the measurement chain at the two cavity
frequencies. (c) Output noise data (colorscale) as a function of detuning
from the cavity frequencies and loop phase. Lines indicate the cuts
shown in (a). (d) Fit of the data in (c) to a coupled-mode theory
with the mechanical environment occupation numbers as free parameters.}
\label{fig:noise-1} 
\end{figure}

An ideal isolator for applications to signal processing and quantum
information would be both efficient and noiseless. To characterize
the noise properties of the device while the four drives are on, we
measure the noise spectrum at the cavity outputs. In Fig.~\ref{fig:noise-1}(a),
we show the signal flow diagrams corresponding to the ideal scattering
matrix (Eq.~\ref{eq:Smatrix}) at the two optimal loop phases. Importantly,
the power input to the mechanical modes (namely, thermal noise) should
appear at the isolated cavity but not the other cavity. The measured
noise spectra shown in Fig.~\ref{fig:noise-1}(b) demonstrate this
behavior. At the loop phase that isolates cavity 1 from cavity 2 (near
$-38$~degrees in green), a noise peak of about 7 photons appears
at cavity 1. The behavior reverses at the opposite loop phase. Data
as a function of frequency and loop phase are shown in Fig.~\ref{fig:noise-1}(c),
with horizontal lines indicating the cuts used in Fig.~\ref{fig:noise-1}(b).

We fit the noise spectra to our expanded model using the parameters
determined from the driven response fit as fixed inputs (Fig.~\ref{fig:noise-1}d).
The only remaining free parameters are the thermal occupation numbers
of the two mechanical environments, $n_{1}$ and $n_{2}$. Equation~\eqref{eq:Smatrix}
predicts the output noise of the isolated port to be the average of
these two occupation numbers. In our system, off-resonant interactions
naturally damp and cool the mechanical modes, yielding lower effective
occupation numbers of the environment $n_{j,\text{eff}}=\Gamma_{j}n_{j}/\Gamma_{j,\text{eff}}$,
measured to be $n_{1,\text{eff}}=0.89\pm0.09$ and $n_{2,\text{eff}}=12\pm1$.
The occupancies of the mechanical modes themselves depend on the loop
phase, with their maxima and minima occurring at $\phi=0$ and $\phi=180$~degrees,
respectively. From the fit to the data in Fig.~\ref{fig:noise-1},
we infer that these mechanical occupancies range from 0.13 to 0.60
phonons in mode one and from 1.5 to 3.7 phonons in mode two. Future
implementations of the optomechanical isolator could reduce the output
noise by starting with lower effective mechanical environment occupation
numbers, for example by introducing additional beam-splitter interactions
to further damp and cool the mechanical modes outside the nonreciprocal
loop.

The device reported here represents a significant advancement of nonreciprocal
technology using optomechanical resources. We have derived closed-form
expressions for the optimal drive conditions required for ideal isolation
and have experimentally implemented them in a microwave optomechanical
circuit. We have fully characterized the nonreciprocal performance
of the device, both in the scattering parameters and the output noise.
The ability to reach high optomechanical cooperativity combined with
the use of an expanded coupled-mode model to fit the data and tune
parameters has allowed us to improve upon crucial metrics of isolation,
approaching the stringent requirements of quantum information processing.
In addition, the quantitative agreement between data and theory shown
here will be crucial for further optimizing performance within experimental
constraints as well as developing more complex multimode systems. 

Looking forward, the scheme we have employed can be straightforwardly
applied to other optomechanical systems, including those at optical
frequencies. The addition of optomechanical systems to the nonreciprocal
parametric toolbox offers the new possibility to directionally route
acoustic signals, and could enable nonreciprocal microwave-to-optical
transduction. Because the theory of the device applies generally beyond
optomechanical systems, the nonreciprocal behavior described here
could also be explored in other parametric systems including microwave
resonators coupled through Josephson junctions. Parametric nonreciprocity
is a promising and quickly developing field, which may soon enable
previously unattainable efficiencies for both measurement and control
of classical and quantum systems.

\paragraph{Note}

While preparing this manuscript, we became aware of another work using
a similar method to demonstrate optomechanical nonreciprocity \cite{BernierTothKoottandavidaEtAl2016}.

\emph{Official contribution of the National Institute of Standards
and Technology; not subject to copyright in the United States.}

\section{Supplementary Information}

\subsection{General theory of a four-mode isolator}

We will use the language and notation established in Ref.~\cite{LecocqRanzani2017}
to analyze a four-mode isolator. We characterize each mode, regardless
of its physical manifestation, by a natural frequency $\omega_{j}$,
a linewidth $\gamma_{j}$, and a signal frequency $\omega_{j}^{s}$
set by the drive frequencies and weak probe input frequency. Modes
$j$ and $k$ can be coupled with a complex coupling rate $g_{jk}$.
We describe the four-mode loop configuration by a mode-coupling matrix,
\begin{equation}
\mathbf{M}=\left(\begin{array}{cccc}
\Delta_{1} & 0 & \beta_{13} & \beta_{14}\\
0 & \Delta_{2} & \beta_{23} & \beta_{24}\\
\beta_{13}^{*} & \beta_{23}^{*} & \Delta_{3} & 0\\
\beta_{14}^{*} & \beta_{24}^{*} & 0 & \Delta_{4}
\end{array}\right),
\end{equation}
where $\Delta_{j}=(\omega_{j}^{s}-\omega_{j})/\gamma_{j}+i/2$ is
the normalized complex detuning of mode $j$, and $\beta_{jk}=g_{jk}/\sqrt{\gamma_{j}\gamma_{k}}$
is the normalized complex coupling strength between modes $j$ and
$k$.%
\footnote{Note that the definition of $\beta$ differs from that in Ref.~\cite{LecocqRanzani2017}
by a factor of two to coincide with the conventional definition of
$g$ in the optomechanics literature.%
} In our system, modes 1 and 2 are microwave cavities and modes 3 and
4 are mechanical. The normalized magnitude of susceptibility for mode
$j$ plotted in Fig.~\ref{fig:diagram} is $1/|\Delta_{j}|^{2}$.
To clarify the analytic results, we assume $|\beta_{13}|=|\beta_{23}|\equiv\beta_{3}$
and $|\beta_{14}|=|\beta_{24}|\equiv\beta_{4}$; that is, each mechanical
mode is equally coupled to both cavity modes. We also put an explicit
$e^{i\phi}$ on \textbf{$\beta_{14}$} for the loop phase, so that
the mode-coupling matrix becomes 
\begin{equation}
\mathbf{M}=\left(\begin{array}{cccc}
\Delta_{1} & 0 & \beta_{3} & \beta_{4}e^{i\phi}\\
0 & \Delta_{2} & \beta_{3} & \beta_{4}\\
\beta_{3} & \beta_{3} & \Delta_{3} & 0\\
\beta_{4}e^{-i\phi} & \beta_{4} & 0 & \Delta_{4}
\end{array}\right).
\end{equation}

The scattering matrix is found from $\mathbf{S}=i\mathbf{HM}^{-1}\mathbf{H}-\mathbf{1}$,
where $H_{jk}=\delta_{jk}\sqrt{\eta_{j}}$. We require nonreciprocity
to occur at the cavity resonance frequencies. This demand lets us
set $\Delta_{1}=\Delta_{2}=i/2$. On resonance, the real parts of
the mechanical detunings are equal to the detunings of the drives
from the red sidebands: $\Delta_{3,4}=\delta_{3,4}+i/2$, where $\delta_{j}$
is a normalized detuning such that the drive frequency is $\omega_{jk}=\omega_{j}-\omega_{k}+\gamma_{k}\delta_{k},$
for $j\in\{1,2\}$ and $k\in\{3,4\}$.

We first require the device impedance matching ($S_{11}=S_{22}=0$)
on resonance. In the high-cooperativity and $\eta_{j}=1$ limits,
impedance matching results in the condition $\delta_{3}=-\delta_{4}$
and gives the optimal detuning as 
\begin{equation}
\delta_{3,\text{opt}}=-\delta_{4,\text{opt}}=\pm\frac{1}{2}\sqrt{2C_{3}C_{4}(1-\cos\phi)-1},
\end{equation}
where $C_{j}=4\beta_{j}^{2}$ is the cooperativity associated with
the optomechanical interaction involving mode $j\in\{3,4\}$.

We parameterize isolation in the system by the transmission difference
$\Delta T=|S_{21}|^{2}-|S_{12}|^{2}$. At the optimal drive detuning,
\begin{equation}
\Delta T=\frac{4\eta_{1}\text{\ensuremath{\eta_{2}}}\sin\phi\sqrt{2C_{3}C_{4}(1-\cos\phi)-1}}{2+(1-\cos\phi)\left(C_{3}^{2}+C_{4}^{2}+2C_{3}+2C_{4}-2C_{3}C_{4}\cos\phi\right)}.
\end{equation}
Maximizing transmission difference over phase, we find the optimal
loop phase $\phi_{\text{opt}}=\arccos(1-1/\sqrt{C_{3}C_{4}})$, with
which the transmission difference becomes
\begin{equation}
\Delta T=\eta_{1}\text{\ensuremath{\eta_{2}}}\frac{8\sqrt{C_{3}C_{4}}-4}{\left(C_{3}-C_{4}\right)^{2}+2\left(\sqrt{C_{3}}+\sqrt{C_{4}}\right)^{2}}.
\end{equation}
At high cooperativity, maximizing this function yields $C_{3}=C_{4}\equiv C$
with corrections at order $1/C$, simplifying the transmission difference
to $\Delta T=\eta_{1}\text{\ensuremath{\eta_{2}}}(1-(2C)^{-1})$.
With these conditions applied, the scattering matrix at high cooperativity
and $\eta_{j}=1$ becomes
\begin{equation}
|\mathbf{S}|^{2}=\left(\begin{array}{cccc}
0 & 0 & 1/2 & 1/2\\
1 & 0 & 0 & 0\\
0 & 1/2 & 1/4 & 1/4\\
0 & 1/2 & 1/4 & 1/4
\end{array}\right).\label{eq:Smat}
\end{equation}
Choosing the opposite phase transposes the above matrix. Note that,
while the qualitative behavior remains the same, at low cooperativities
the optimizations discussed above are not exact. In analyzing the
experimental data, we therefore use numerical maximization of the
exact difference function $\Delta T$ to decide how to update the
drive parameters.

To find the bandwidth of nonreciprocity, we calculate the transmission
difference as a function of the detuning $\delta\omega$ from the
cavity centers with the approximation that the cavity widths are much
larger than the mechanical widths. With the above optimizations for
drive detunings, loop phase, and cooperativities, the result is
\begin{equation}
\Delta T(\omega)=\eta_{1}\eta_{2}\frac{\gamma^{*2}}{\gamma^{*2}+4(\delta\omega)^{2}}+\mathcal{O}\left(C^{-\frac{1}{2}}\right),
\end{equation}
where $\gamma^{*}=4\gamma_{3}\gamma_{4}/(\gamma_{3}+\gamma_{4})$.
The above shows that in the high cooperativity limit the bandwidth
of nonreciprocity is independent of cooperativity and equal to 
\begin{equation}
\Gamma_{NR}=4\frac{\gamma_{3}\gamma_{4}}{\gamma_{3}+\gamma_{4}}.
\end{equation}

\subsection{Off-resonant damping and expanded coupled-mode theory}

Due to the off-resonant coupling terms discussed in the text, each
mechanical mode can respond to all four drives. To be able to predict
the effect of changing the drive powers and frequencies, these extra
interactions must be included in the model. Expanding our mode basis
allows us to fit the experimental data using the intrinsic mechanical
properties and also predict the needed drive parameters to obtain
optimal performance.

\begin{figure}
\includegraphics[width=1\columnwidth]{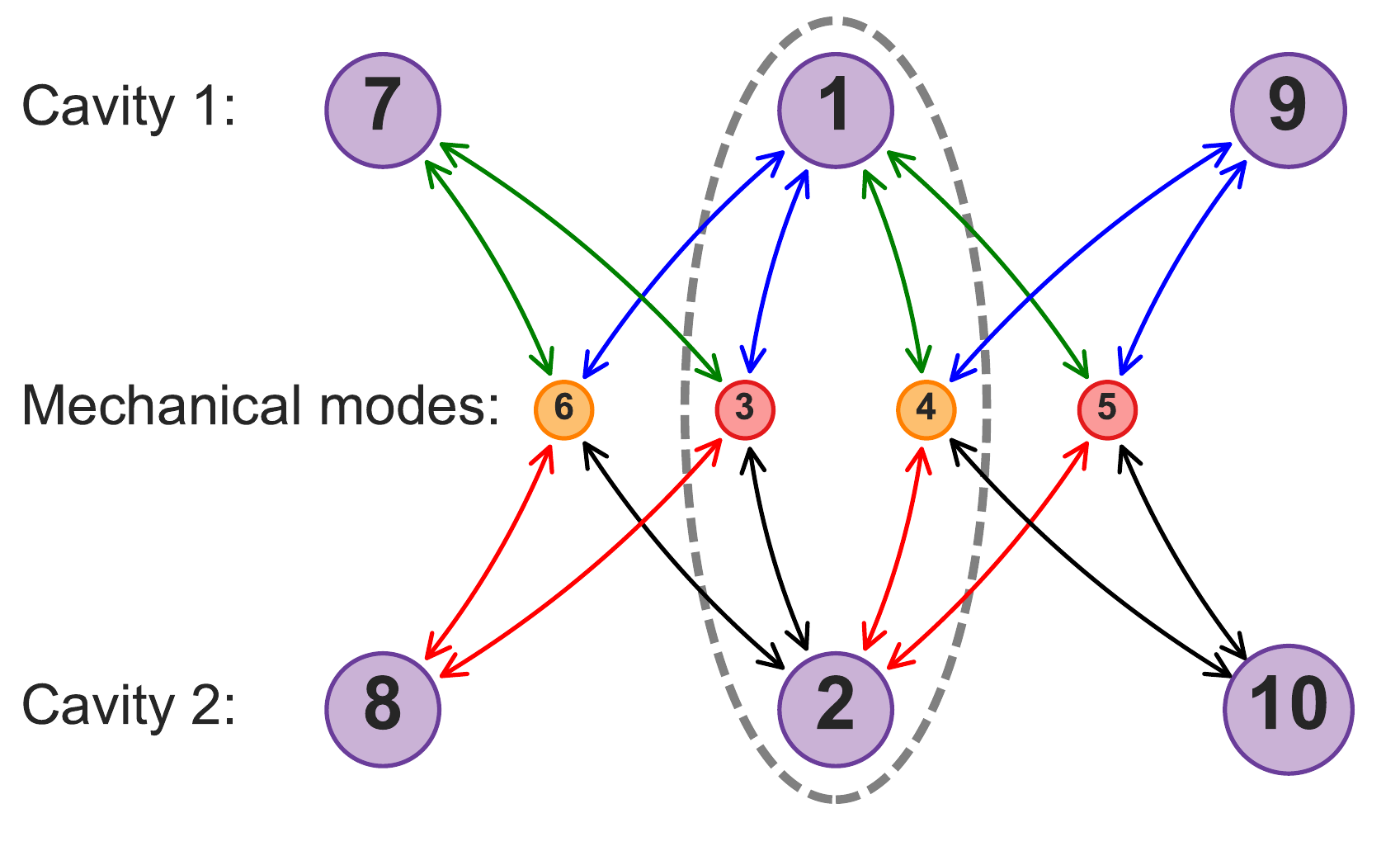}

\protect\caption{Ten-mode graph diagram. Like-colored double-sided arrows indicate
optomechanical coupling driven by the same microwave drive. Modes
1, 2, 3, and 4 are the four modes in the simplified four-mode model.
Modes 5 through 10 are duplicates evaluated off resonance.}

\label{fig:10modeDiagram}
\end{figure}

The expanded mode basis needed, diagrammed in Fig.~\ref{fig:10modeDiagram},
comes directly from the coupled equations of motion. In the diagram,
like-colored arrows indicate interactions driven by the same microwave
drive. Modes 1 through 4 are the four modes appearing in the simplified
four-mode model discussed above. Modes 5 through 10 are auxiliary
modes evaluated at the relevant off-resonant frequencies determined
by the drives. For example, the signal frequency of mode 7 is $\omega_{7}^{s}=\omega_{1}^{s}-\omega_{13}+\omega_{14}$,
while that of mode 8 is $\omega_{8}^{s}=\omega_{2}^{s}-\omega_{23}+\omega_{24}$.
As our analysis takes place in the Fourier domain, each of these distinct
coupled frequencies acts as another mode, even if it resides in the
same physical oscillator as another mode. For this reason, modes 1,
7, and 9 share the resonance frequency and linewidth of cavity 1.
Likewise for modes 2, 8, and 10 in cavity 2, and for the mechanical
mode pairs \{3, 5\} and \{4, 6\}.

A note is needed to justify the presence of the off-resonant mechanical
modes 5 and 6. In general, these extra modes are needed to maintain
common linewidths and frequencies of all the auxiliary cavity modes.
This effect is typically negligible in optomechanics because the cavities
are so much wider than the mechanical modes. Another reason for including
modes 5 and 6, however, is to be able to model the scattering parameters
over wide spans that include both resonant and off-resonant structure.
We therefore include the off-resonant mechanical terms to able to
fit wide scans of scattering parameters.

In total, these considerations lead to our system of ten modes that
quantitatively accounts for the off-resonant damping. Notably, we
ignore all amplification processes occurring at the blue sidebands.
This is a reasonable approximation because the damping effects from
these terms are smaller by a factor of $\kappa^{2}/(16\Omega^{2})<1\%$.

We have justified the need for an expanded model and shown how to
find the signal frequencies of the modes. The last part needed before
calculating the scattering matrix are the couplings involving the
auxiliary modes. These are found by relating all 16 couplings to the
4 original couplings by multiplying by ratios of vacuum optomechanical
coupling rates and intrinsic mechanical linewidths.

\begin{figure*}
\includegraphics[width=1\textwidth]{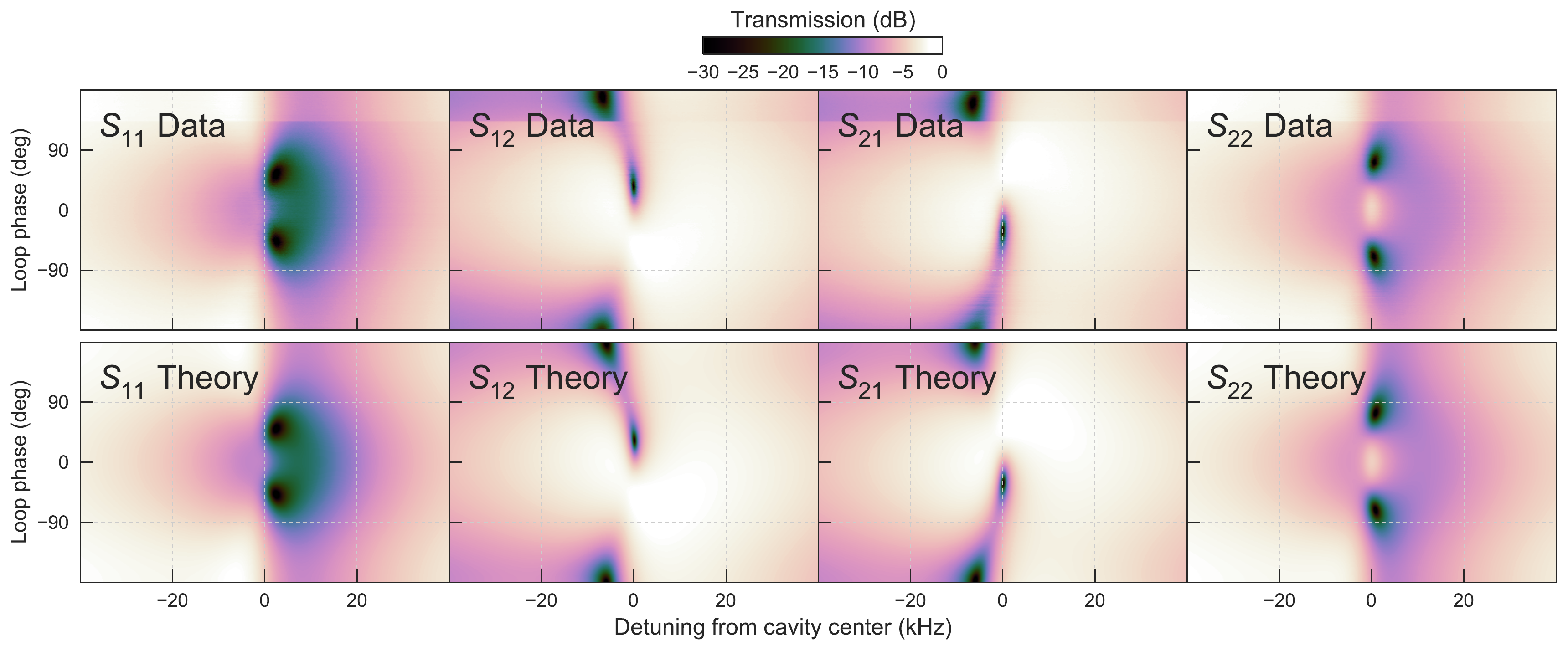}\protect\caption{Full scattering matrix including transmission and reflection. Measured
scattering parameters are shown on the top row and the results of
a least-squares fit to our expanded coupled-mode theory are shown
on the bottom row.}
\label{fig:fullSfit}
\end{figure*}

With the mode-coupling matrix fully determined, we proceed to calculate
the scattering matrix as above. We use this expanded model for the
scattering parameters to fit the data shown in the text and to predict
the needed drive parameters to maximize the transmission difference
function. Figure~\ref{fig:fullSfit} shows the full fit including
the reflection coefficients.

The ten-mode graph can be reduced to obtain an effective four-mode
graph. By allowing the inputs for the auxiliary modes to be exactly
zero, one can derive the effective mode-coupling matrix describing
the reduced system. This reduction procedure is a classical approximation,
so care must taken in its application to quantum noise calculations.
In general, to reduce mode $k$ from the matrix $\mathbf{M}$, we
perform the transformation
\begin{equation}
M_{ij}^{\prime}=M_{ij}-\frac{M_{ik}M_{kj}}{M_{kk}},
\end{equation}
which results in a new matrix $\mathbf{M}^{\prime}$ with one less
dimension. Reducing each auxiliary mode in turn results in the effective
four-mode model. Incidentally, the mode reduction formula encodes
the meaning of the rotating wave approximation in Fourier space; if
the correction to element $M_{ij}$ is negligible for all signal frequencies
of interest, the dynamics of mode $k$ can be safely ignored.

\subsection{Calculation and calibration of output noise}

Here we calculate a model for the output noise given the ($10\times10$)
scatting matrix calculated above. We start with the system output
amplitude, then model the amplifier chain and the spectrum analyzer.

The output amplitude for mode $j$ in terms of the $N$ input amplitudes
is
\begin{equation}
\hat{a}_{j\text{,out}}=\sum_{k=1}^{N}S_{jk}\hat{a}_{k\text{,in}}.
\end{equation}
The output amplitude is then amplified, which we model as a transformation
to another mode operator, $\hat{c}_{j}$, by
\begin{eqnarray}
\hat{c}_{j} & = & \sqrt{G_{j}}\hat{a}_{j,\text{out}}+\sqrt{G_{j}-1}\hat{d}_{j}^{\dagger},
\end{eqnarray}
where $G_{j}$ is the gain at port $j$, and $\hat{d}_{j}^{\dagger}$
is an input creation operator used to model the amplifier's added
noise. When the mode $\hat{c}_{j}$ is fed into the spectrum analyzer,
the measured noise power spectrum $\mathcal{N}[\omega]$ is \cite{ClerkDevoretGirvinEtAl2010}
\begin{equation}
\mathcal{N}_{j}[\omega]=\hbar\omega\int_{-\infty}^{\infty}\frac{d\omega^{\prime}}{2\pi}\left\langle \hat{c}_{j}^{\dagger}[\omega]\hat{c}_{j}[\omega^{\prime}]\right\rangle .
\end{equation}
Taking the large gain limit (so that $G_{j}-1\simeq G_{j}$), and
using input correlators $\left\langle \hat{a}_{j,\text{in}}^{\dagger}[\omega]\hat{a}_{j,\text{in}}[\omega^{\prime}]\right\rangle =2\pi n\delta(\omega-\omega^{\prime})$
for a thermal state with occupancy $n$, we find
\begin{equation}
\mathcal{N}_{j}[\omega]=\hbar\omega G_{j}\left(1+n_{j,\text{amp}}+\sum_{k=1}^{N}\left|S_{jk}\right|^{2}n_{k,\text{th}}\right),
\end{equation}
where $n_{j,\text{amp}}\ge0$ is the noise from the amplifier, and
$n_{k,\text{th}}\geq0$ is the thermal occupation number for the input
field at port $k$. We measure $\mathcal{N}_{j}[\omega]$ in units
of W Hz$^{-1}$. Knowing the system gain and added noise allows us
to convert the spectrum to units of output photons from the device.
When the set of $n_{k,\text{th}}$ (and possibly the $n_{j,\text{amp}}$)
are the only fit parameters, the model is linear and can therefore
be fit to the data using linear least-squares fitting methods. The
third term in the above equation is what we refer to as the output
noise of the device. We measure the amplification noise at the two
cavity frequencies to be $n_{1,\text{amp}}=30\pm3$ and $n_{2,\text{amp}}=22\pm2$.

We calibrate the output noise by heating the cryostat to 100 mK and
measuring single drive optomechanical spectra \cite{TeufelDonnerLiEtAl2011}.
This process yields the system gain, system added noise, and the four
vacuum optomechanical coupling rates, which are found for each cavity-mechanical
mode pair to be $|(g_{11}^{(0)},g_{12}^{(0)},g_{21}^{(0)},g_{22}^{(0)})|/2\pi\simeq(50,40,60,20)$~Hz,
where $g_{jk}^{(0)}$ is the vacuum coupling rate for the $j$th cavity
and the $k$th mechanical mode.

\bibliographystyle{hunsrt}
\bibliography{OM_Isolator_bibliography}

\end{document}